\documentclass[amsmath,amsfonts,groupedaddress,notitlepage,twocolumn,nofootinbib,floatfix]{revtex4-1}

\usepackage{lipsum} 
\usepackage[usenames,dvipsnames]{color} 
\usepackage{graphicx}
\usepackage{hyperref}
\usepackage[normalem]{ulem}
\usepackage{cancel}

\definecolor{dark-red}{rgb}{0.4,0.15,0.15}
\definecolor{dark-green}{rgb}{0,0.65,0}
\definecolor{dark-blue}{rgb}{0,0,0.75}

\hypersetup{
    colorlinks, linkcolor={dark-red},
    citecolor={dark-green}, urlcolor={dark-blue},
    breaklinks = true
}

\newcommand{\ee}{{\mathbf e}}
\newcommand{\rr}{{\mathbf r}}

\newcommand{\qq}{{\mathbf q}}
\newcommand{\nn}{{\mathbf n}}
\newcommand{\vv}{{\mathbf v}}
\newcommand{\hh}{{\mathbf h}}

\newcommand{\0}{{\mathbf 0}}
\newcommand{\su}{{\cal U}}
\newcommand{\sv}{{\cal V}}

\newcommand{\ve}{{\mathbf \varepsilon}}

\begin{document}


\title{Phase transitions at high energy vindicate negative microcanonical temperature}
\author{P. Buonsante}
\affiliation{QSTAR \& CNR - Istituto Nazionale di Ottica, 
 Largo Enrico Fermi 2, I-50125 Firenze, Italy.}
\author{R. Franzosi}
\affiliation{QSTAR \& CNR - Istituto Nazionale di Ottica, 
 Largo Enrico Fermi 2, I-50125 Firenze, Italy.}
\author{A. Smerzi}
\affiliation{QSTAR \& CNR - Istituto Nazionale di Ottica, 
 Largo Enrico Fermi 2, I-50125 Firenze, Italy.}



\date{\today}

\begin{abstract}
The notion of negative absolute temperature emerges naturally from Boltzmann's definition of ``surface'' microcanonical entropy in isolated systems with a bounded energy density. Recently, the well-posedness of such construct has been challenged, on account that only the Gibbs ``volume'' entropy ---and the strictly positive temperature thereof--- would give rise to a consistent thermodynamics.
Here we present analytical and numerical evidence that Boltzmann microcanonical entropy provides a consistent thermometry for both signs of the temperature. In particular, we show that Boltzmann (negative) temperature allows the description of phase transitions occurring at high energy densities, at variance with Gibbs temperature. 

Our results apply to nonlinear lattice models standardly employed to describe the propagation of light in arrays of coupled waveguides and the dynamics of ultracold gases trapped in optical lattices. 
Optically induced photonic lattices, characterized by saturable nonlinearity, are particularly appealing because they offer the possibility of observing states ---and phase transitions--- at both signs of the temperature.

\end{abstract}
\maketitle

\section{Introduction}
Since the pioneering work of Purcell, Pound and Ramsey on nuclear spin systems \cite{Ramsey_PR_81_278,*Purcell_PR_81_279,Ramsey_PR_103_20}, negative absolute temperature has been an established concept in Statistical Physics  \cite{Landau_SP1,Kittel_TP}.  The ever growing control of ultracold atoms recently allowed the preparation of negative-temperature states for motional degrees of freedom of a bosonic gas loaded in an optical lattice  \cite{Braun_Science_339_52}.  Despite this remarkable result, the very notion of negative temperature has been challenged in a recent article  \cite{Dunkel_NaturePhysics_10_67}, 
on account that the Boltzmann ``surface'' entropy it stems from would be inconsistent, and the only consistent picture would be based on the Gibbs ``volume'' entropy. This criticism spurred a lively and ongoing debate  \cite{Sokolov_NaturePhysics_10_7,Frenkel_AJP_83_163,Dunkel_arXiv_1403_6058,Vilar_JCP_140_201101,Treumann_arXiv_1406_6639,Schneider_arXiv_1407_4127,Hilbert_PRE_90_062116,Dunkel_arXiv_1408_5392,Treumann_FrontPhys_2_49,Swendsen_arXiv_1410_4619,Campisi_PRE_91_052147,Ferrari_arXiv_1501.04566,Cerino_PC, Wang_arXiv_1507_02022,Hanngi_arXiv_1507_05713,Poulter_arXiv_1508_00350}.
 
Here we address the consistency of Boltzmann microcanonical temperature, focusing on a class of nonlinear lattice models standardly employed in the description of the propagation of light through arrays of waveguides, and the dynamics of ultracold bosons trapped in optical lattices. 
Our most interesting results apply to the case of optically induced photonic crystals \cite{Efremidis_PRE_66_046602,Fleischer_Nature_422_147,Fleischer_PRL_90_023902,Melvin_PRL_97_124101}. We find that, at variance with standard nonlinearity, the saturable nonlinearity characterizing these models {\it supports both positive-- and negative-temperature states in the same system}. Even more interestingly, we show that the same physical system can undergo phase transitions for critical energies both in the lower and in the upper portion of the (bounded) energy spectrum. A finite-size scaling analysis shows that the Boltzmann picture provides a consistent description in both cases. While the former correspond to the standard situation, the critical temperature of the latter turns out to be {\em finite and negative}. Despite they typically manifest themselves in a clear ordering of the system, phase transitions at high energy densities are not consistently captured by the Gibbs picture. More in general, the Gibbs temperature does not appear to be a measurable quantity for the systems under concern.

As to the standard ``cubic'' nonlinearity  typical of ultracold lattice bosons, we confirm that it does support negative-temperature states, as proposed in Refs.~\cite{Mosk_PRL_95_040403,Rapp_PRL_105_220405} and experimentally demonstrated in Refs.~\cite{Braun_Science_339_52,Braun_PNAS_112_3641}. Owing to a pathological scaling of the energy density, 
states with positive or negative temperature turn out to be unstable for attractive or repulsive interactions, respectively. Nevertheless, phase transitions at negative temperatures should be observable in the former case, as heralded by the ordering phenomena observed experimentally \cite{Braun_Science_339_52,Braun_PNAS_112_3641}. 


We support our conclusions with analytic arguments and extensive (microcanonical) numerical simulations. In particular,  we provide independent and concordant tests of the fact that the considered dynamical states correspond to thermal equilibrium. We measure their (Boltzmann) temperature ---which can be either positive or negative depending on the (conserved) energy density--- either as a (time-averaged) function of the instantaneous configuration of the dynamical variables \cite{Franzosi_JSP_143_824} or through a fit of the average distribution of the relevant modes of the system. 
Also, we show how, irrespective of the sign of the temperature, a large lattice acts as a thermostat for a small sublattice, thus confirming the equivalence between isolated and thermostated systems that is crucial for a consistent definition of temperature. From a different point of view,
 being able to support negative temperature states, the sublattice can be in principle used as a thermometer for the whole lattice. 

The plan of this paper is the following. We briefly survey the concept of negative (Boltzmann) temperature in Sec.~\ref{NTS}, and introduce the nonlinear lattice models we focus on in Sec.~\ref{MODS}. After briefly addressing ensemble equivalence in Sec.~\ref{ENSS}, we  discuss phase transitions ---at both positive and negative temperatures--- in Sec.~\ref{PTS}. The results of our numerical simulations are presented in Sec.~\ref{TTS}. More detail about our results can be found in the appendix sections. 

\section{Negative Absolute Temperatures}
\label{NTS}
In the microcanonical ensemble, the inverse temperature of the system is defined as 
\begin{equation}
\label{betas}
\beta = \frac{1}{k_{\rm B}}\frac{\partial s}{\partial h}
\end{equation}
where $s(h)$ is the entropy density corresponding to the energy density $h$, and  $k_{\rm B}$ is Boltzmann's constant.
In general, two choices for $s(h)$ are possible, corresponding to Boltzmann's and Gibbs' definitions.
According to the former, $s(h) =V^{-1} k_{\rm B} \log\big(\omega(h)\, \Delta h\big) $, where $V$ is the number of degrees of freedom in the system, $\omega(h)$ is the density of microstates at energy density $h$ and $\Delta h$ is a constant having the same dimension as $h$.   
The Gibbs entropy is obtained by replacing $\omega(h)$ with $\Omega(h) = \int_{h'<h} dh' \,\omega(h')$, i.e. the number of microstates having an energy density up to the chosen one. As it is well known, in the thermodynamic limit these two definitions are equivalent in ``standard'' systems lacking an upper bound to the energy density, for simple geometrical reasons (see e.g. \cite{Reichl_aMCiSP}). This comes about because $\omega(h)$ is an increasing function of $h$. However, some systems exist where the energy density has an upper bound and $\omega(h)$ is  a (non-negative) concave function featuring a maximum at some finite energy density $h_*$. 
The logarithm of $\omega(h)$ clearly has the same properties, which entails that $\beta(h)$ is positive for $h<h_*$,   negative for $h>h_*$, and vanishes at $h=h_*$. A simple version of the lattice models introduced in Sec.~\ref{MODS} is employed to exemplify this behavior in Appendix~\ref{BEA}.

The key ingredient for the occurrence of negative Boltzmann temperatures is the existence of an upper bound to the available energy densities. Furthermore, the elements of the thermodynamical system must be in equilibrium, so that a temperature can be consistently defined. Finally, the system must be thermally isolated from any system that do not meet the previous requirements \cite{Ramsey_PR_103_20}. 
In most cases the first condition fails because of the unboundedness of the kinetic energy term.
The first experiments demonstrating negative Boltzmann temperatures  \cite{Ramsey_PR_81_278,*Purcell_PR_81_279} involved spin systems, which are not affected by this problem owing to the lack of kinetic degrees of freedom. 
The kinetic energy density of a gas of particles can be effectively bounded from both above and below in the presence of a periodic potential inducing an energy gap sufficiently large that the physics of the system is dominated by states in the lowest energy band. 
 Building on this observation, negative-temperature states for motional degrees of freedom in an ultracold bosonic gas have been demonstrated in a recent experiment \cite{Braun_Science_339_52}.

However, as observed in Ref~\cite{Dunkel_NaturePhysics_10_67}, $\omega(h)\geq 0$ ensures that $\Omega(h)$ is a non-decreasing function of $h$, and hence the  Gibbs temperature is non negative even in systems with a bounded energy density. Therefore, plugging the Boltzmann or Gibbs entropy in Eq.~\eqref{betas} can produce very different temperatures, and the question arises as to which picture is the correct one.

The appearance of a negative sign in an absolute temperature might look disturbing enough to discard the Boltzmann framework at first glance. It should be noted, however, that  a state at negative temperature is not colder than ``the coldest possible state'', i.e. the ground state. Since its energy density exceeds $h_*$, it is in fact hotter than the state attaining the maximum Boltzmann entropy, which has an infinite temperature. 
The fact that the concept of  ``hotter that $T=\infty$'' sounds still somewhat disturbing  can be merely ascribed to the traditional use of $T=(k_{\rm B}\beta)^{-1}$ in the scale of temperatures. Using $-\beta$ instead of $T$ restores the ``correct order'' of cold and hot in the whole range of Boltzmann temperatures\footnote{Here we do not introduce a minus sign in front of $\beta$, in order to avoid confusion.} \cite{Ramsey_PR_103_20}.


In the following, we produce analytical and numerical evidence that the Boltzmann entropy does in fact provide a consistent thermodynamic picture.
As we mention in the introduction, Ref.~\cite{Dunkel_NaturePhysics_10_67} advocates that only the Gibbs entropy results in a consistent thermostatics, and dismisses all previous claims about negative absolute temperatures. These arguments are further elaborated in Refs.~\cite{Hilbert_PRE_90_062116,Hanngi_arXiv_1507_05713}.   While we refer the Reader to a different publication \cite{confutation} for a more systematic discussion of the points raised in Refs.~\cite{Dunkel_NaturePhysics_10_67,Hilbert_PRE_90_062116,Hanngi_arXiv_1507_05713}, in the following we  occasionally comment on some of them.

We start by observing that disturbing features also lurk behind the instinctively appealing 
non-negative temperatures characterizing the Gibbs formalism. Indeed, for lattice systems such as the ones we are going to address shortly, the Gibbs temperature corresponding to energy densities $h\geq h_*$ increases arbitrarily with the number of degrees of freedom in the system. Hence, in the thermodynamic limit, it is  {\em identically infinite} on the whole finite interval of energy densities exceeding the one attaining the maximum Boltzmann entropy\footnote{This is  the main criticisms leveled by Ref.~\cite{Vilar_JCP_140_201101} agains the Gibbs picture advocated by Ref.~\cite{Dunkel_NaturePhysics_10_67}. We note that this weird feature of the Gibbs temperature is mentioned in Ref.~\cite{Dunkel_NaturePhysics_10_67} itself, albeit only in the Supplementary Information. }. Correspondingly, the Gibbs heat capacity would be  {\em identically zero}. We observe that, while this might be to some extent internally consistent, it presents at least two problems, as we illustrate in more detail in Sec.~\ref{PTS} and Appendix~\ref{BEA}. In the first place, it clearly makes the Gibbs temperature incapable of describing phenomena involving states with $h>h_*$. Also, the Gibbs temperature cannot be measured as a microcanonical  average by exploiting the usual formulation of the equipartition theorem advocated in Refs.~\cite{Dunkel_NaturePhysics_10_67,Hilbert_PRE_90_062116}.

We remark that the convexity properties of the density of states $\omega(h)$ of the systems under investigation are a typical consequence of the large number of degrees of freedom and short-range interactions. The oscillating density of states discussed in Refs.~\cite{Hilbert_PRE_90_062116,Hanngi_arXiv_1507_05713} can crop up in systems with a small number of degrees of freedom or long-range interactions.  Standard thermodynamic relations can be ---at least formally--- used in such cases, although at the expense of features that appear crucial for a sensible definition of temperature \cite{Cerino_PC}. For instance, the equivalence of isolated and thermostated systems is lost. Also, when joined, two systems having the same temperature could equilibrate to a completely different temperature. The well-posedness and usefulness of the concept of temperature in such situations is at least arguable.

We finally note that some of the arguments against the existence of negative-temperature equilibrium states are based on the failure of one or more of the key conditions listed above \cite{Ramsey_PR_103_20}. For instance, it has been argued that such states would not be stable if the system in which they occur is brought into contact with a system that is unable to sustain negative-temperature states, e.g.  due to the lack of an upper bound to the energy density \cite{Dunkel_NaturePhysics_10_67,Hilbert_PRE_90_062116,RomeroRochin_PRE_88_022144} . This, however, is basically a truism, since the composite system clearly does not meet the requirements \cite{Ramsey_PR_103_20} for sustaining negative-temperature equilibrium states.

The model we are going to address in the following has no doubt been devised as an extreme idealization  of real physical systems, as remarked in Refs.~\cite{Hilbert_PRE_90_062116,Hanngi_arXiv_1507_05713}. On the other hand, over the last few years enormous steps towards the faithful {\it experimental simulation} of similarly ideal models have been made, the trailblazer being ultracold-atom  physics \cite{Jaksch_AnnPhys_315_52,Bloch_NatPhys_8_267}. Lattice Hamiltonians originally conceived as rough yet extremely challenging toy models have been experimentally realized with an high degree of fidelity by loading a cold atomic gas into a ``crystal of light''. As we mention, evidence of negative-temperature states has already been reported for these systems \cite{Braun_Science_339_52}. The observation of phase transitions occurring at negative critical temperature, such as the ones we discuss in Secs.~\ref{PTS} and \ref{TTS} involves a similar effort towards the faithful experimental simulation of the single-band lattice model described in the following Section.  
On a related note, we mention that a  Mott insulator-superfluid {\it quantum} phase transition has been observed at negative temperature in the experimental
realization of a Bose-Hubbard model with attractive interactions~\cite{Braun_PNAS_112_3641}. We once again remark that significant steps in the realization of synthetic nonlinear lattice models have been likewise made in the field of optical waveguides. These include  an analysis of the effects of nonlinearity the Anderson localization \cite{Lahini_PRL_100_013906} and the observation of a Berezinskii-Kosterlitz-Thouless phase transition at positive critical temperature \cite{Situ_FIO_2012} in optical systems obeying the discrete nonlinear Schr\"odinger (DNLS) equation.    

\section{The model}
\label{MODS}
We focus on nonlinear lattice models \cite{Kevrekidis_DNLSE} of the form
\begin{equation}
\label{H}
H = U \sum_{\rr} u(|z_\rr|^2)-J \sum_{\rr \rr'} z_\rr^* A_{\rr \rr'} z_{\rr'}
\end{equation}
where $\rr = (r_1,r_2,\cdots,r_d)$ denotes a site in a $d-$dimensional ($d$D) lattice  and $A_{\rr \rr'}$ is the relevant coordination matrix. The coordinates of the sites are integer numbers, $r_j = 1,2,\cdots,L_j$, so that the total number of sites in the lattice is $V=\prod_{j=1}^d L_j$. Periodic boundary conditions are assumed, i.e. $z_{\rr +L_j \ee_j} = z_\rr$, where $\ee_j$ is the versor along the $j$-th direction. We set the hopping amplitude to $J=1$, so that the units for energy and time are $J$ and $\hbar J^{-1}$, respectively. As to the nonlinear term, we consider two cases
\begin{equation}
\label{u12}
 u_{1}(n) = -\log(1+n),\qquad u_{2}(n) = \frac{1}{2} n^2.
\end{equation}
The former corresponds to the saturable nonlinearity typical of the equations describing the propagation of a light probe in an optically induced nonlinear photonic lattice \cite{Efremidis_PRE_66_046602,Fleischer_Nature_422_147,Fleischer_PRL_90_023902,Melvin_PRL_97_124101}, while the latter produces the cubic nonlinearity of standard DNLS equations. These are employed in the description of diverse phenomena \cite{Kevrekidis_DNLSE,Lederer_PhysRep_463_1}, including the 
  dynamics of ultracold atoms loaded in optical lattices \cite{Cataliotti_Science_293_843,Trombettoni_PRL_86_2353,Smerzi_PRL_89_170402,Cataliotti_NJP_5_71,Trombettoni_NJP_7_57,Flach_PhysRep_467_1,Pikovsky_PRL_100_094101,Kopidiakis_PRL_100_084103,Flach_PRL_102_024101,Rumpf_PRE_69_016618,Small_PRA_83_013806,Iubini_NJP_15_023032,Franzosi_Nonlinearity_24_R89}
and the propagation of light in waveguide arrays \cite{Morandotti_PRL_83_2726,*Morandotti_PRL_86_3296,Rumpf_PRE_69_016618,Small_PRA_83_013806,Iubini_NJP_15_023032,Franzosi_Nonlinearity_24_R89,Lederer_PhysRep_463_1,Christodoulides_Nature_424_817,Lahini_PRL_100_013906,*Lahini_PRL_103_013901}. 

The equations of motion generated by Hamiltonian~\eqref{H} via the Poisson brackets $\{z_j,z_\ell^*\} = -i \hbar^{-1}\delta_{j \ell}$ have two first integrals, the energy and ``particle'' density, 
\begin{equation}
h = V^{-1}H, \qquad a = V^{-1}\sum_\rr |z_\rr|^2.
\end{equation} 
The presence of a conserved quantity other than the energy density is important for the occurrence of negative temperatures, because it makes the configuration space of any finite system compact.

In the non interacting limit $U\to0$ the Hamiltonian becomes linear, and the (thermo)dynamics is described exactly by the single-particle ``plane-wave" eigenmodes $z_\rr^{(\qq)} = \sqrt{a} \,e^{i (\qq \cdot \rr - \ve_\qq t)}$, where $\frac{L_j}{2\pi} q_j = 0,\,1,\,2,\cdots L_j-1$ is the quasimomentum along direction $j$. The corresponding single-particle energies $\ve_\qq =  -2   \sum_{j=1}^d \cos q_j$ form a band bounded by $\pm 2 d $. 

It is easy to check that the ``plane-wave'' states  are normal modes for the nonlinear equations as well, provided that the single-particle energy is replaced by the frequency $\nu_\qq(a) =  U u'(a)+  \ve_\qq $. The corresponding energy density is $h_\qq(a) =  U u(a)+ a\, \ve_\qq$. For repulsive interactions, $U>0$ ---i.e. {\em defocusing} nonlinearity--- the energy densities are bounded from below by $h_{\0}(a) = U u(a)-2 d a$. On finite lattices the energy densities also have an upper-bound, which however diverges in the thermodynamic limit for the standard nonlinearity, $u_2$. Note indeed that the energy density of a state where only an individual site is occupied is $V^{-1} u_2(a V) = U/2 a^2\, V$. This means that negative-temperature equilibrium states are problematic for the standard defocusing nonlinearity~\cite{Rumpf_PRE_69_016618}, although metastable states at $\beta<0$ can persist for astronomically long times on 1D lattices \cite{Iubini_NJP_15_023032,Franzosi_Nonlinearity_24_R89}. 

In the case of saturable nonlinearity, $u_1$, the upper bound of the energy density remains finite in the thermodynamic limit, and tends to $h_{\rm max} = a \ve_{\mathbf \pi} = 2 d a $. That is, the energy {\em per particle} is of the order of the maximum single-particle energy.

The situation for attractive interactions $U<0$, ---i.e. for {\em self-focusing} nonlinearity--- is related to the previous case through the mapping
\begin{equation}
\label{map}
H\left(U,J,\{ z_\rr\}\right)\! =\!-H\left(-U,J,\{e^{i \pi \sigma_\rr} z_\rr\}\right),\; \sigma_\rr\!= \sum_{j=1}^d r_j
\end{equation}
This means that  negative temperatures are well defined for the standard nonlinearity, $u_2$, and positive ones are problematic.  Note that switching the interaction strength to negative values is a crucial step for obtaining negative-temperature states in a bosonic gas loaded in an optical lattice~\cite{Braun_Science_339_52}. The nonlinearity is typically self-focusing also in the case of waveguide arrays. Although the sign of $U$ can be reversed in photorefractive crystals  \cite{Efremidis_PRE_66_046602,Christodoulides_Nature_424_817}, the defocusing case does not seem to lend itself to a tight-binding approach in current experimental realizations \cite{Efremidis_PRE_66_046602}.  
For these reasons, unless otherwise specified, in the following we fix our attention mainly on the {\em self-focusing} case, $U<0$. 

\section{Ensembles}
\label{ENSS}
In general, the details of an experimental system determine the most appropriate choice for the statistical ensemble to be adopted  in the description of its thermodynamic properties. The natural choice for an isolated system is the microcanonical ensemble. For ergodic systems, one expects that microcanonical thermodynamic quantities can be equivalently obtained as ensemble or temporal averages. 

In the canonical and grand canonical ensembles, $\beta$ is a Lagrange multiplier fixing the total energy. It is possible to prove that in the thermodynamic limit this multiplier coincides with the microcanonical definition of temperature, Eq.~\eqref{betas} \cite{confutation}.
Also, canonical ---or, in the presence of additional conserved quantities, grand canonical--- time averages can be obtained by considering  a sufficiently macroscopic subset of an isolated, microcanonical system. In the absence of pathologies, one expects that the subsystem has the same thermodynamical properties as the whole system. More in general, one expects that all ensembles provide a consistent description of a given system.

In the non-interacting limit, $U=0$, a detailed analysis of the statistical ensembles for the model in Eq.~\eqref{H} can be carried out, at both the semiclassical and quantum level. In both cases, the relation between the energy density and the inverse temperature turns out to be the same for all the  three ensembles \cite{confutation}. The easiest way of obtaining such relation is through the grand-canonical ensemble. The grand partition function for the model in Eq.~\eqref{H} is
\begin{equation}
\label{ZGC}
{\cal Q} = \int \prod_\rr dz_\rr e^{-\beta V\left[h(\{z_\rr\})-\mu a(\{z_\rr\})\right]}, 
\end{equation} 
where $\mu$ is the chemical potential, i.e. the Lagrange multiplier selecting the average density $a$, and we omitted the dependance of the energy density on the parameters. In the non-interacting limit, the integral in Eq.~\eqref{ZGC} can be easily carried out. It is likewise easy to obtain the average occupation of the single particle modes\footnote{\label{classicaln}This is the ``classical version'' of the Bose-Einstein distribution $n_\qq = [e^{\beta(\ve_\qq-\mu)}-1]^{-1}$, that comes about because the occupation of the single-particle modes of Eq.~\eqref{H} is not restricted to integers.}
\begin{equation}
\label{nq}
n_\qq(\beta,\mu) = \langle |{\tilde z}_\qq|^2 \rangle = \frac{1}{\beta} \,\frac{1}{\ve_\qq -\mu},
\end{equation}
where ${\tilde z}_\qq = V^{-1} \sum_\rr e^{i \rr\cdot\qq} z_\rr$ is the Fourier transform of the configuration of the system.
For fixed $\beta$, $a$, and $h = a \kappa$  ---where $\kappa$ denotes the kinetic energy density per particle---, the chemical potential $\mu(\beta,a,\kappa)$ can be found by inverting the relations
\begin{equation}
\label{ah}
a = \sum_\qq n_\qq,\qquad h = \sum_\qq \ve_\qq n_\qq.
\end{equation}
On a sufficiently large one-dimensional lattice this calculation can be carried out analytically, and gives
\begin{equation}
\label{betamu1D}
\beta = -\frac{1}{a}\frac{2 \kappa}{4-\kappa^2}, \qquad \mu = \frac{\kappa^2+4}{2\kappa}.
\end{equation}
Thus, for an equilibrium thermodynamic state, $\beta>0$  if $-2<\kappa<0$, and $\beta<0$ if $0<\kappa<2$.  

As we mention, the first of Eqs.~\eqref{betamu1D} accurately describes the relations between $\beta$, and $\kappa$ and $a$ that are found in the canonical and microcanonical ensembles \cite{confutation}. This means that, in the limit of a large number of sites $L$, $\omega(h,a) \sim [4-(h/a)^2]^L$ . Using the Laplace method it is possible to evaluate $\Omega(h,a)$, and the Gibbs entropy thereof. As illustrated in Appendix~\ref{BEA}, it turns that the Gibbs inverse temperature is the same as in the Boltzmann picture in the lower energy interval, and vanishes identically on the whole upper interval. This causes the failure of the standard equipartition theorem, which underlies the measurability of the Gibbs temperature as a microcanonical average (see Sec.~\ref{TTS} ).  

The thermodynamics of model\footnote{Our choice for the Poisson brackets $\{z_{\rr^{}}^{},\,z_{\rr'}^*\}$ corresponds to the standard bosonic commutation rules when the C-number $z_\rr$ is interpreted as the expectation value of an on-site boson operator, e.g. in the Bose Hubbard model. Refs~\cite{Rasmussen_PRL_84_3740,Samuelsen_PRE_87_044901} make a different choice for the same Poisson brackets. The two choices are connected by a simple mapping. We illustrate the results of Refs~\cite{Rasmussen_PRL_84_3740,Samuelsen_PRE_87_044901} in the light of our choice.}  \eqref{H} on one-dimensional lattices has been addressed in Refs.~\cite{Rasmussen_PRL_84_3740} and~\cite{Samuelsen_PRE_87_044901}  for {\em defocusing} standard and saturable nonlinearity, respectively. There, the grand-canonical partition function, Eq.~\eqref{ZGC} is calculated using a transfer-matrix approach,
which allows the identification of the the region in the $(a,h)$ plane corresponding to positive temperatures. This is bounded from below by the ground-state energy, $h^{(\infty)} = U u_m(a)-2 a$, and from above by a critical line $h_m^{(0)}(a)$ that, in the case of standard nonlinearity, assumes the simple form $h_2^{(0)}(a) = 2 U u_2(a) = U a^2$ (the superscript in the energy densities here refers to the grand canonical inverse temperature).  In Ref.~\cite{Rasmussen_PRL_84_3740} the region  $h>h_2^{(0)}(a)$ is argued to correspond to negative temperatures, based on the change in the concavity of the probability distribution function of the amplitudes $|z_\rr|^2$, as obtained from microcanonical dynamical simulations. Ref.~\cite{Samuelsen_PRE_87_044901} repeats basically the same analysis as in Ref.~\cite{Rasmussen_PRL_84_3740}, but the only new insight it provides about the region $h>h^{(2)}(a)$ is the observation that initial states picked in that region  end up having one single very mobile localized excitation. This is contrasted with the larger number of pinned localized excitations characterizing the standard nonlinearity \cite{Rumpf_PRE_69_016618,Iubini_NJP_15_023032,Rasmussen_PRL_84_3740}. 

\begin{figure}
\begin{tabular}{cc}
 \includegraphics[height=3.6cm]{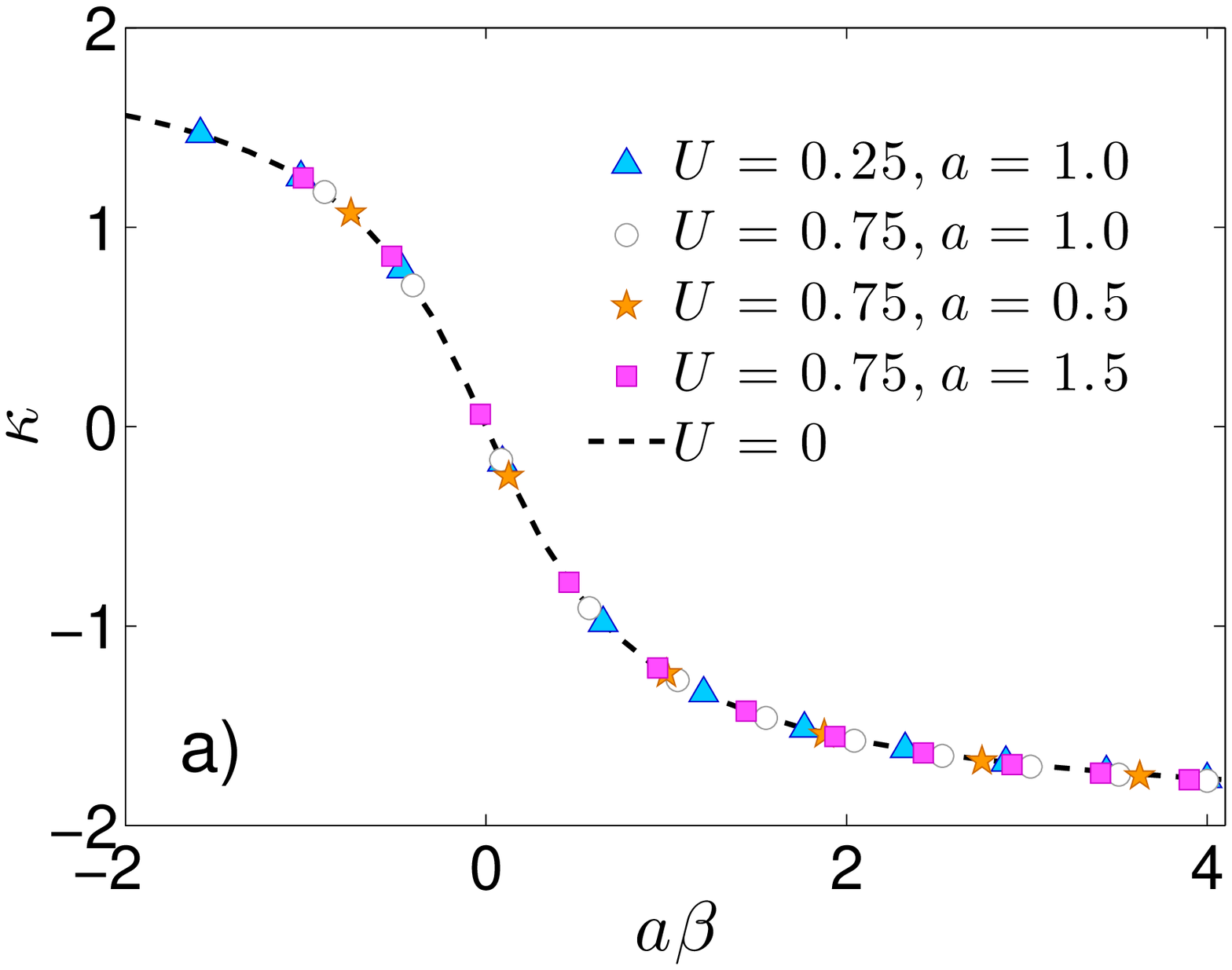} &   \includegraphics[height=3.6cm]{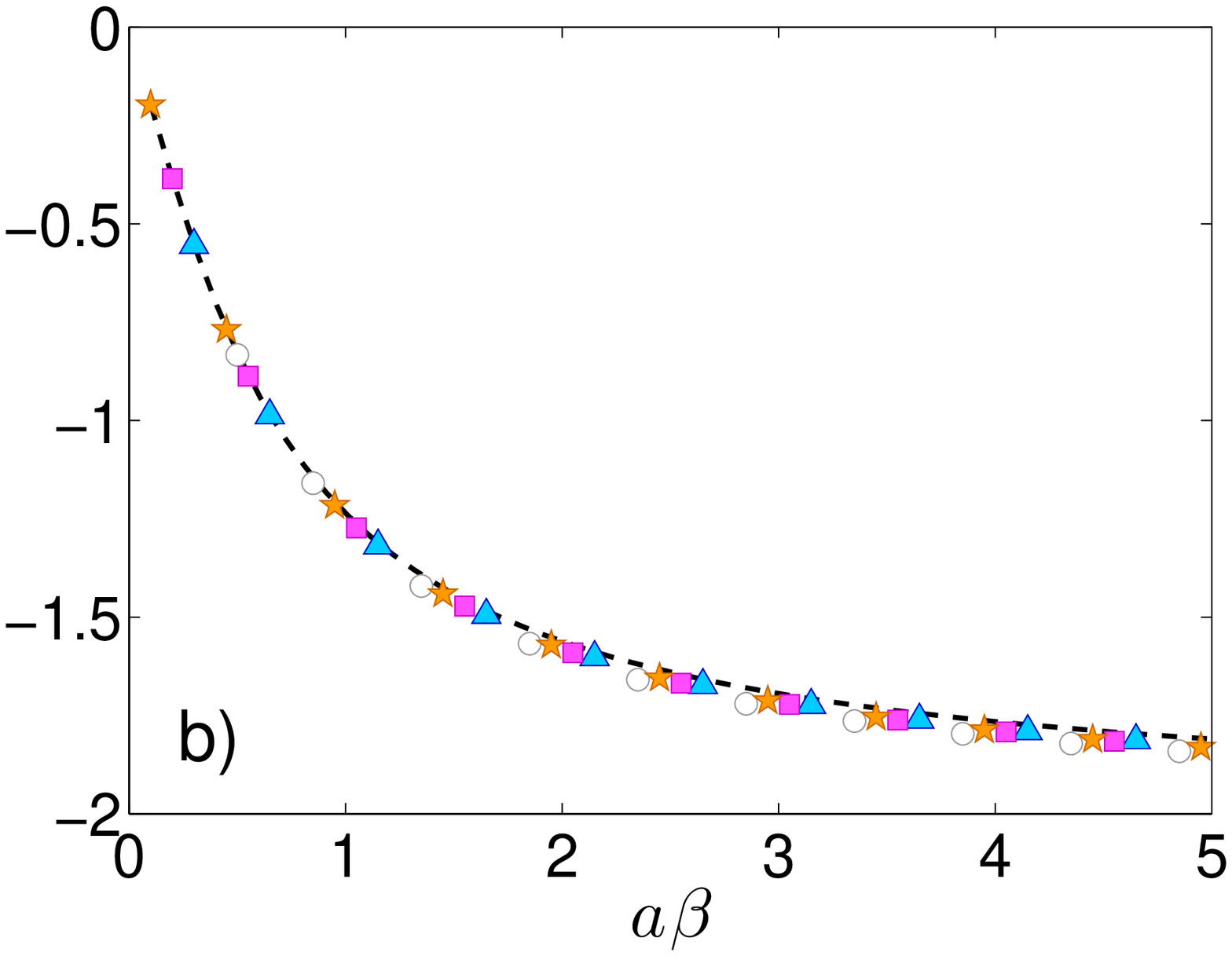} 
 \end{tabular}
\caption{\label{transmat} Relation between the grand-canonical $\beta$ and the {\em kinetic} energy per particle in a 1D lattice,  as provided by the transfer-matrix approach (symbols). Panels a) and b) refer to saturable and standard  nonlinearities, respectively. Despite the non-negligible interaction strengths, the data points closely follow the corresponding non-interacting result, Eq.~\eqref{betamu1D}. }
\end{figure}

We find that, in view of the boundedness of the available energy densities, the transfer-matrix approach  applies also for $\beta<0$ for the defocusing saturable nonlinearity considered in Ref.~\cite{Samuelsen_PRE_87_044901}.
Specifically it can be applied for  $\beta>\beta_{\rm L}$, where $\beta_{\rm L}<0$ in general depends on $U$ and $a$. 
This is illustrated in Fig.~\ref{transmat}, where we analyze the relation between between $\beta$ and the {\em kinetic} energy per particle $\kappa$ in the interacting case, as provided by the transfer matrix approach. In particular, it is clear from panel  a), that solutions exist at negative $\beta$ for the saturable nonlinearity. The leftmost symbol of each kind marks the largest negative $\beta$ we were able to analyze for the corresponding parameter choice. The failure of the transfer-matrix approach for larger negative $\beta$'s is related to a phase transition between an extended and a localized state, occurring at a finite negative $\beta$. Indeed, as we discuss in the following, states with\footnote{We have verified that the energy thresholds $h_m^{(\infty)}(a)$ \cite{Rasmussen_PRL_84_3740,Samuelsen_PRE_87_044901} also apply to two and three dimensional lattices \cite{extended}. } $h>h_2^{(\infty)}(a)$ exhibit a persistent and mobile localized excitation only for sufficiently large energies, i.e. for sufficiently small negative temperatures \cite{extended}. Panel b) in Fig.~\ref{transmat} illustrates similar results for the case of standard nonlinearities, where the transfer matrix approach clearly fails as soon as $\beta<0$ \cite{Rasmussen_PRL_84_3740}.
It is interesting to note that, despite the non-negligible effective interaction, the data points closely follow the analytical relation derived in the non-interacting case, Eq.~\eqref{betamu1D}. Our microcanonical simulations confirm that this is the case also for two-- and three--dimensional lattices \cite{extended}. 
We once again remark that the above described situation is reversed for self-focusing nonlinearities. 

\section{Phase Transitions}
\label{PTS}
 In the non-interacting limit, the model in Eq.~\eqref{H} is known to undergo Bose-Einstein condensation on a three-dimensional lattice. If $\beta>\beta_{\rm C}>0$, that is if the energy density is sufficiently small,  a macroscopic fraction of the particle density occupies the ground state of the system, i.e. the kinetic-energy eigenstate corresponding to quasimomentum $\qq = (0,0,0)$. On a finite-size lattice this second-order phase transition manifests itself as a crossover. For this transition to occur, it is crucial that the density of states in the vicinity of the ground state has a suitable behavior.
 Since this behavior is literally mirrored by the density of states in the vicinity of the highest-energy state\footnote{We remark that this symmetry in the system spectrum and in the relevant Boltzmann entropy does not imply the physical or thermodynamical equivalence of  states having opposite energy density, as argued in Refs. \cite{Dunkel_arXiv_1408_5392,Hilbert_PRE_90_062116}. In fact, for one, the states are distinguished by the sign of the derivative of the Boltzmann entropy, and of the temperature thereof.}, it seems  fair to expect that the system condenses into the highest-energy  eigenstate for sufficiently large energy densities ---i.e. at small negative temperatures---. Specifically, one expects that for  $\beta<-\beta_{\rm C}$ a macroscopic density of particles occupies the state $\qq = (\pi,\pi,\pi)$. In fact, this is  what results from a simple grand-canonical calculation (see Fig.~\ref{BEC3DU0} in Appendix \ref{BECGC}). We mention that  clear signatures of phase transitions at high-energies have been recently discussed for short-range ferromagnets \cite{Kastner_PRL_102_240604}. 
 Also, the emergence of order at small negative temperature in an isolated planar superfluid has been recently discussed in Ref. \cite{Simula_PRL_113_165302}, thus confirming an early prediction by L. Onsager \cite{Onsager_NC_6_279}.

The condensation transition occurring at large $\beta>0$ is expected to survive the introduction of a defocusing standard nonlinear term, $u_2$ (see e.g. Ref.~\cite{Ramakumar_PRB_72_094301}). As we observed earlier, such nonlinearity has a dramatic effect on high-energy states. The upper bound to the energy density, and the negative temperatures thereof, are lost in the thermodynamic limit.  In view of the mapping in Eq.~\eqref{map}, a condensation into the highest-energy state is expected at a negative critical temperature in the case of self-focusing standard nonlinearity.

The saturable nonlinearity, $u_1$,  produces less dramatic effects. It is not hard to check that for $U<0$ 
the ground state is localized. Specifically, the corresponding particle density features a single peak  of finite width (corresponding to a breather), on top of a uniform background\footnote{This self-trapped state can be centered at any lattice site, and hence it is not unique. This simmetry breaking is a well known feature in nonlinear systems. }. The highest-energy state is instead extended, and coincides with the uniform ``plane-wave'' state with $\qq=(\pi,\pi,\pi)$.
 It is therefore tempting to envisage two phase transitions for this system: a condensation into the extended highest-energy state for $\beta<\beta_{\rm E} < 0$, and a {\it self-trapping transition} ---i.e. a ``condensation'' into the localized ground-state--- for $\beta>\beta_{\rm L} > 0$. In the defocusing case the localization properties of the  extremal states are swapped, and hence one expects  a condensation into an extended ground-state for $\beta>\beta_{\rm E'} = -\beta_{\rm E}$, and a condensation into a localized highest-energy state for $\beta<\beta_{\rm L'} =-\beta_{\rm L}$. According to the Mermin-Wagner theorem, condensation into an uniform state is not expected to occur for $d\leq 2$ at a finite critical temperature, since it involves the breaking of the continuous symmetry in the phases of the dynamical variables $z_\rr$. 
The localized ground-state instead breaks the {\em discrete} translational symmetry of the lattice, and does not exhibit long-range order. Therefore, the corresponding localization transition is not excluded for $d<3$.

Also, the model in Eq.~\eqref{H} is strictly related to the classical XY model, and it is therefore expected  to undergo a Berezinskii-Kosterlitz-Thouless (BKT) transition at a finite temperature on a 2D lattice \cite{Trombettoni_NJP_7_57,Small_PRA_83_013806}.
Thus a particularly intriguing scenario opens up for two-dimensional optically induced nonlinear photonic lattices, which realize model~\eqref{H} for self-focusing saturable nonlinearity \cite{Fleischer_Nature_422_147,Efremidis_PRE_66_046602}. One could observe a localization transition at finite positive temperatures, {\em and} a BKT transition at finite negative temperatures. In fact, signatures of the latter transition have been reported for defocusing nonlinearity at positive temperatures \cite{Situ_FIO_2012}.


\section{Thermalization and Thermometry}
\label{TTS}
The numerical integration of the dynamical equations generated by Hamiltonian~\eqref{H} reveals that, after a possibly long transient, the system reaches a stationary state in which the instantaneous value of observables characterizing the system performs small oscillations about an asymptotic value (see Appendix~\ref{TTA}). 
The observables we typically consider are for instance the kinetic and interaction energy per particle,
\begin{align}
\kappa &= -\frac{1}{V a}\sum_{\rr,\, \rr'} z_\rr^* A_{\rr,\, \rr'} z_{\rr'} =\frac{1}{a} \sum_\qq |{\tilde z}_\qq|^2 \ve_\qq \\
 {\cal I} & = a^{-1} h- \kappa = \frac{U}{V a} \sum_\rr u(|z_\rr|^2)
\end{align}
where ${\tilde z}_\qq$ denotes the Fourier transform of $z_\rr$.
More importantly and interestingly, it is possible to give an estimate of the instantaneous microcanonical Boltzmann temperature as a function of the instantaneous configuration of the system \cite{Rugh_PRE_64_055101,Davis_PRA_66_053618,Franzosi_JSP_143_824} (see Appendix \ref{TTA} for details).
A time average excluding the initial transient,
\begin{equation}
\label{tavg}
\langle {\cal O} \rangle = \frac{1}{\Delta t} \int_{t_0}^{t_0+\Delta t} dt'\, {\cal O}\left(\{z_\rr(t')\}\right) 
\end{equation} 
provides a ``measure'' of the generic observable $\cal O$, where we stress once again that this includes the Boltzmann temperature. Plotting $\langle \kappa \rangle$ vs $a \langle \beta \rangle$ reveals that the relation between these quantities remains remarkably close to the one applying in the non-interacting limit also for non-negligible nonlinearity, although some small deviations appear for the symmetry-broken phases \cite{extended}.


An additional evidence of thermalization is provided by the observation that the prediction in Eq.~\eqref{nq} is fulfilled  by the ``relevant modes'' in the system, which we generically denote $\zeta_\qq$. 
 That is, a plot of $\langle |\zeta_\qq|^2 \rangle^{-1}$ vs. the corresponding energies $\epsilon_\qq$ results into a straight line whose slope coincides with the time-averaged measure of the  microcanonical Boltzmann temperature $\langle \beta \rangle$. In the absence of condensation the relevant modes are the single particle modes, i.e $\zeta_\qq = {\tilde z}_\qq$ and $\epsilon_\qq = \ve_\qq$. In the condensed phase the linear relation is fulfilled by the Bogoliubov quasiparticle modes (see Appendix \ref{TTA} for details. Figures~\ref{betaC} and \ref{BogM} contain examples of this behavior). Therefore, the interactions not only drive the system to equilibrium but, when this is reached, maintain it by acting as a ``heat bath'' for the relevant, effectively non-interacting modes of the system. One could argue that the above slope represents a sort of canonical or (even grand-canonical) measure of the temperature, since, while the total population of the  modes may be strictly conserved (in the case of the kinetic modes), the relevant total energy is not. A grand-canonical description is also obtained by considering only the dynamical variables $z_\rr$ belonging to a sufficiently large sublattice of the whole lattice. Indeed, the first integrals of the motion, $h$ and $a$, are not conserved when restricted to a portion of the whole sample. The fact that the relevant modes of the sublattice behave as those in the whole lattice (as apparent in Fig.~\ref{betaC} in Appendix~\ref{TTA}),  is a further proof that the system is in equilibrium, and that the Boltzmann temperature has the properties expected of a temperature.
If the whole lattice is much larger than the sublattice, then the former acts as a thermostat (and a ``chemostat'') for the latter.
We also checked that a grand-canonical Langevin approach generalizing the one introduced in Ref.~\cite{Iubini_JSM_P08017},  where $\beta$ is an external parameter, produces results in agreement with the above described observations for both positive and negative temperatures \cite{extended}. 

A further fundamental test for a well defined temperature concerns the equilibration of two systems that are  separately at equilibrium at different temperatures. One expects that, when these are brought into contact, energy ---and, in our case, particles--- flows in agreement with the intuitive notion of cold and hot, in such a way that eventually the inverse temperature of the composite system is intermediate between the two initial values.
 As discussed in Ref.~\cite{confutation}, this is exactly the case for the Boltzmann temperature, irrespective of the sign it initially has in the separate systems. Note that some care must be taken when using a small system as a thermometer for a larger system. Indeed,  both energy and particles will be in general exchanged  to attain equilibrium \cite{confutation,extended}.
 Of course, in a thermalization experiment where two systems with opposite signs of the temperature are brought into contact, the resulting composite system should support both positive-- and negative--temperature states. For instance, if one of the two initially separate systems only supports positive temperatures and the other supports both, there is no chance that the equilibrium state of the composite system be negative, irrespective of the initial temperature of the latter subsystem \cite{Ramsey_PR_103_20}.  In this respect, in view of its ability to support both positive-- and negative--temperature states, a nonlinear lattice system characterized by saturable nonlinearity seems to be an ideal setting for this kind of experiments, at least in principle. 
 
\begin{figure}
\begin{tabular}{cc}
\includegraphics[width=4.2cm]{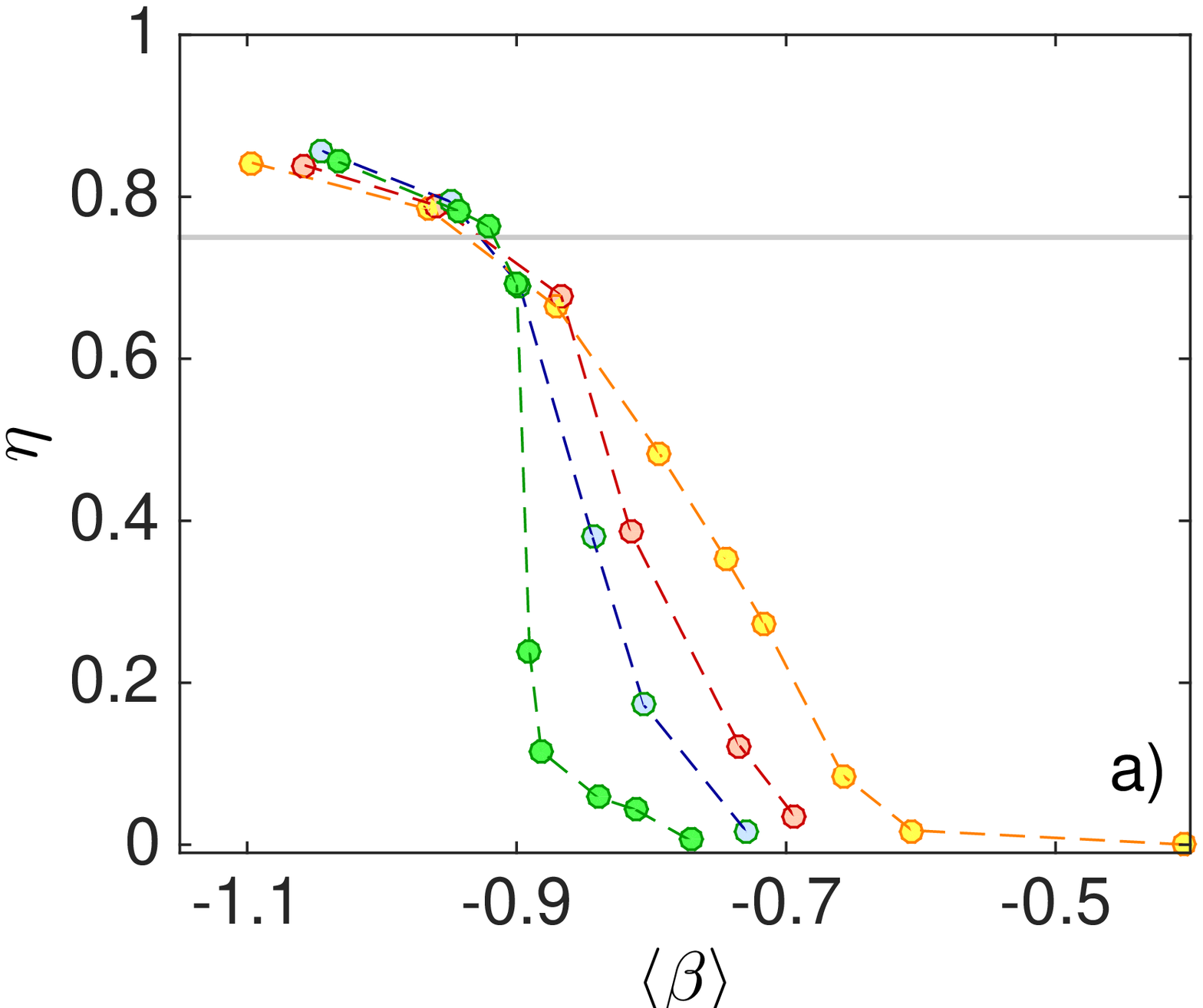} & \includegraphics[width=4.2cm]{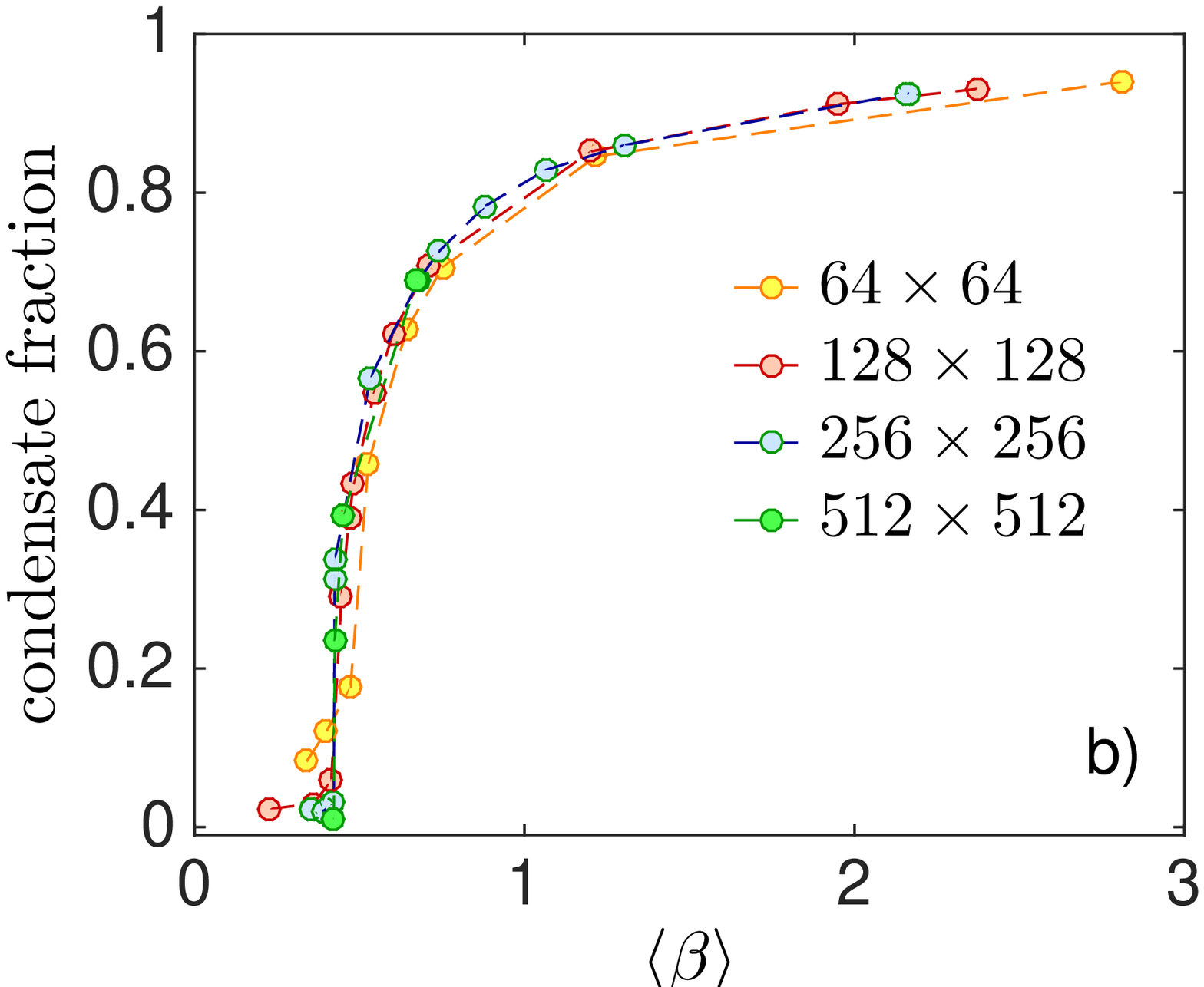}
\end{tabular}
\caption{\label{2Dsat} Phase transition in a 2D lattice model with saturable self-focusing nonlinearity ($a=1$, $U=-0.75$); a) BKT transition at negative Boltzmann temperature. The horizontal gray line signals the expected critical point; b) condensation into the localized ground-state at positive temperature. The dashed lines are guides to the eye.}
\end{figure}

\begin{figure}
\includegraphics[width=4.5cm]{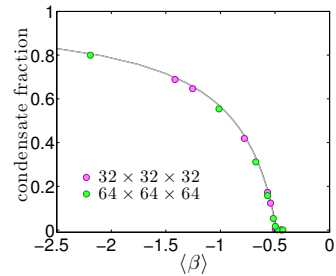} 
\caption{\label{cond3DC} Condensation transition at negative Boltzmann temperature in a 3D lattice model with standard self-focusing nonlinearity ($U=-0.75$, $a = 0.5$). The solid line is the prediction from the Bogoliubov approximation.}
\end{figure}

  The descriptive power of the Boltzmann microcanonical temperature becomes  fully evident in the presence of phase transitions, as demonstrated in Figs. \ref{2Dsat} and \ref{cond3DC}.  Figure~\ref{2Dsat}  refers to the 2D lattice system with self-focusing saturable nonlinearity modeling the propagation of light through an optically induced photonic lattice  \cite{Efremidis_PRE_66_046602,Fleischer_Nature_422_147,Fleischer_PRL_90_023902}. As we anticipated in Sec.~\ref{PTS}, two phase transitions occur in such a system. A suitably defined exponent $\eta$, sensitive to the decay properties of the radial correlations,  signals a BKT transition in the system \cite{Small_PRA_83_013806} (see Appendix~\ref{aBKT}). 
The data we obtain  for different lattice sizes shown in Fig.~\ref{2Dsat} a), strongly suggest a transition at negative critical temperature, and cross at the value $\eta = 3/4$ expected at the critical point \cite{Small_PRA_83_013806,Polkovnikov_PNAS_103_6125} (see Appendix~\ref{aBKT} for more detail). As we observed earlier, for sufficiently large $\beta>0$ the microcanonical state of the system develops a density peak. This suggests that the system is partially condensed into its ground-state, which is also characterized by a (taller) density peak. We quantify this condensation through the time-averaged projection of the microcanonical state of the system onto the ground-state\footnote{The occurrence of a density peak breaks the (discrete) symmetry of the lattice. Thus, we calculate the overlap of the instantaneous dynamical state and the ground-state after centering the relevant peaks at the same lattice site. }. As illustrated by panel b) of  Fig.~\ref{2Dsat}, a plot of such quantity against the Boltzmann temperature of the corresponding microcanonical state clearly signals the occurrence of a phase transition.

Finally, Fig.~\ref{cond3DC} shows the average occupation of the highest-energy state in a three dimensional lattice system with standard self-focusing nonlinearity. 
The data points, obtained as temporal microcanonical averages according to Eq.~\eqref{tavg}, clearly signal a (condensation) phase transition, and nicely follow the (solid gray) curve obtained from a grand-canonical calculation based on the Bogoliubov approximation (see Appendix~\ref{TTA}).
The same behavior is obtained for the saturable nonlinearity on a three dimensional lattice. In the self-focusing case one observes the condensation into the localized ground state at small positive temperatures, and the condensation into  the  extended highest-energy state state for small negative temperatures. Again, the relevant critical temperatures are finite \cite{extended}. 
 
In the same physical situations the limitations of the Gibbs picture become evident. Although the microcanonical dynamics takes place in the phase-space ``sheet'' corresponding to a given value of the energy density (and, possibly, of other conserved quantities), the Gibbs entropy requires information about all energies below such value. It is therefore not immediately clear how the Gibbs (inverse) temperature $\beta_{\rm G}$ could be obtained as a microcanonical ensemble average (and hence a time average). It is often argued \cite{Dunkel_NaturePhysics_10_67,Hilbert_PRE_90_062116} that this is made possible by the ``equipartition theorem''
\begin{equation}
\label{equip}
\beta_{\rm G}^{-1} = \left\langle	\zeta_j \frac{\partial H}{\partial \zeta_j }\right\rangle,
\end{equation}
where $\zeta_j$ denotes any  element  of the set of dynamical variables describing the microstate of the system, and the angle brackets denote the standard microcanonical average. However, Eq.~\eqref{equip} typically fails for system admitting negative Boltzmann temperatures.
For instance, for a negative-temperature state of the self-focusing case of Eq.~\eqref{H}, the r.h.s. of  Eq.~\eqref{equip}  with $\zeta_j=z_j^*$ is of the order of $a[\langle \kappa \rangle + U u'(a)]$. As we have discussed, the corresponding $\beta_{\rm G}$ tends instead to vanish as the system size increases, so that  Eq.~\eqref{equip} cannot be possibly satisfied, since its l.h.s. diverges. This failure  of the ``equipartition theorem'' stems from ignoring a surface term in the derivation of Eq.~\eqref{equip}, which is legitimate only for systems that do not admit negative Boltzmann temperatures \cite{confutation,extended,Cerino_PC} (where the Boltzmann and Gibbs pictures are equivalent, as we discussed before). 
Even if $\beta_{\rm G}(h)$ were actually measurable, its usefulness would be of very limited value in describing the phenomena involving the upper part of the spectrum, such as the phase transitions illustrated in Figs.  \ref{2Dsat} and \ref{cond3DC}. Indeed, while the critical energy density (and Boltzmann temperature) for such transitions becomes size-independent for sufficiently large systems, the corresponding Gibbs temperature indefinitely increases with the system size.

\section{Discussion}
We have addressed the statistical physics of nonlinear lattice models relevant in the description of the propagation of light in nonlinear media and the dynamics of ultracold atoms trapped in optical lattices. We have discussed how the Boltzmann picture provides a consistent description of equilibrium and equilibration processes, and  that the absolute temperature characterizing the equilibrium states can have either sign. Negative temperatures correspond to high energy states, and come about due to the presence of an upper bound to the available energy density, and from the decreasing character of the entropy (density) in the vicinity of such bound. These features are already apparent in the noninteracting limit of the considered lattice models \cite{confutation}, and survive the introduction of interactions, provided that these do not give rise to pathological scaling in the thermodynamic limit. Interactions act as a heat bath for the relevant, effectively non interacting modes of the system, driving the system towards equilibrium. A large system can act as a thermostat for a smaller system, bringing it to a negative-temperature (i.e. higher-energy) state, provided that such a state can be supported by the composite system. Such a  process might not fit the definition of ``conventional heating'' \cite{Dunkel_NaturePhysics_10_67}, but it is consistently described in terms of Boltzmann (inverse) temperature.
A likewise consistent description is instead problematic in the Gibbs picture, where the whole upper  interval of available energy densities corresponds to zero heat capacity and infinite temperature  (which, in addition, is not measurable as a microcanonical average through the standard equipartition theorem).  For the same reason, the description of the phase transitions taking place in the system at high energy densities is similarly problematic. As illustrated in Figs. \ref{2Dsat} and \ref{cond3DC}, Boltzmann temperatures provide a consistent description of such transitions.

Optical systems represent an ideal testbed for our conclusions. Ordering phenomena related to the self-trapping transition discussed above have been observed in one-- \cite{Fleischer_PRL_90_023902} and two-dimensional \cite{Fleischer_Nature_422_147} lattices. Signatures of a BKT transition have been observed in a two-dimensional optically induced photonic lattice \cite{Situ_FIO_2012}, although they have been analyzed in terms of an ``equipartition'' {\it effective} temperature \cite{Small_PRA_83_013806}. Optically induced nonlinear photonic lattices  \cite{Efremidis_PRE_66_046602,Fleischer_Nature_422_147,Fleischer_PRL_90_023902,Situ_FIO_2012} are particularly intriguing, in that   
the relevant saturable nonlinearity preserves both the upper and lower bound characterizing the  corresponding (single-band) linear lattice model, allowing the exploration of both  positive- and negative-temperature states in the same system. Even more interestigly, the realizability of two-dimensional lattices \cite{Fleischer_Nature_422_147,Situ_FIO_2012} opens up the possibility of observing phase transitions with critical temperature of both signs in the same (synthetic) physical system.
 States in the upper portion of the kinetic energy band can be excited by suitably tilting the input beam, as described in Refs.~\cite{Christodoulides_Nature_424_817,Lahini_PRL_100_013906}.
Phase transitions on lower dimensional systems could be engineered through the introduction of suitable ``on-site'' or topological defects \cite{StamperKurn_PRL_81_2194,Burioni_EPL_52_251}. 

The measure of the instantaneous microcanonical temperature requires the knowledge of the instantaneous configuration of the field, $z_\rr(t)$. A more feasible measure is obtained through Eq.~\eqref{nq}. Indeed, as discussed above and  demonstrated in Figs.~\ref{betaC} and \ref{BogM} in Appendix~\ref{TTA},  a linear fit of the inverse average mode occupation versus the corresponding  energy provides an estimate of the temperature that is in remarkable agreement with the time-average of the instantaneous microcanonical value.

\acknowledgments
P.B. gratefully acknowledges helpful suggestions by R. Burioni, A. Vezzani and S. Wimberger.
\appendix
\section{Boltzmann Entropy}
\label{BEA}
\begin{figure}
\includegraphics[width=5cm]{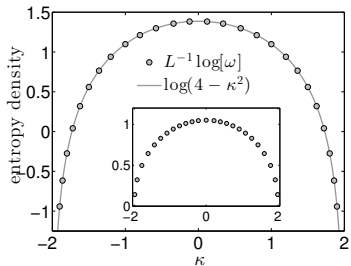}
\caption{\label{BEntropy} Boltzmann entropy density for a 1D noninteracting lattice model comprising $L=20$ sites. Main plot: semiclassical case, Eq.~\eqref{H}. The inset shows the situation for the corresponding quantum (Bose-Hubbard) model. In all cases $a=1$.  }
\end{figure}
Figure~\ref{BEntropy} illustrates the concepts discussed in Sec.~\ref{NTS}, in the case of a uniform 1D lattice model comprising $L$ sites and containing $a L$ non-interacting bosons, ($U=0$, $a=1$, $L=20$). The circles in the main panel correspond to the analytically calculated 
\cite{confutation} volume of the phase space relevant to the chosen energy and particle density
\begin{equation}
\omega(h,a) = \int \prod_\qq dz_\qq \,\delta\left(L h -H \right)\delta\left(L a-N \right)
\end{equation}
where $H$ is the Hamiltonian in Eq.~\eqref{H} and 
\begin{equation}
\label{N}
N = \sum_\qq |z_\qq|^2
\end{equation}
Although the number of sites is not very large, the microcanonical result is very well described by the approximation $\omega \approx C (4-\kappa^2)^L$  derived from the grand canonical result in Eq.~\eqref{betamu1D}, where $\kappa = a^{-1} h$. The data in the inset refer to the Bose-Hubbard model obtained by changing the C-numbers $z_\qq$ and $z_\qq^*$ in Hamiltonian~\eqref{H} into the lattice boson operators ${\hat a}_\qq$ and ${\hat a}_\qq^\dag$, respectively. In view of the lack of interaction, a generic eigenstate of the system is a Fock state  listing the number of bosons occupying each of the single-particle states. The behavior of $\omega$ can be therefore estimated by listing all the Fock states compatible with the chosen  boson population and binning the corresponding energy densities. We  divided the energy density interval $[-2a, 2a]$ into 183 bins but, for graphical reasons, plotted only a few of the corresponding numbers of microstates. The result is remarkably smooth owing to the very large number of Fock states (approximately $6.89\times10^{10}$).

As discussed in Sec.~\ref{NTS}, the Boltzmann entropies are concave functions, and feature a maximum at $h_*=0$. Therefore, according to Eq.~\eqref{betas} the corresponding Boltzmann inverse temperatures are positive for $h<h_*$ and negative for $h>h_*$. Note that, owing to the small particle density, the classical and quantum results are quantitatively different, although qualitatively similar.

As we repeatedly mention, in the thermodynamic limit the Gibbs inverse temperature equals Boltzmann's for $h<h_*$ and is identically zero for energy densities exceeding $h_*$. This can be appreciated by plugging analytic function obtained from the grand canonical picture  into
\begin{equation}
\Omega(h,a)=\int_{-2 a}^h dh'\,\omega(h,a).
\end{equation}
Using the Laplace method, for $\ve$ not too close to 0 we get
\begin{equation}
\Omega(h,a)\approx
\left\{
\begin{array}{ll}
\frac{C a^2}{2|h|} [4-(a^{-1} h)^2]^{L+1}, & h \in [-2 a,\, 0) \\[0.5em]
\sqrt{\pi} \,C\, 2^{2L+1} a, & h \in (0,\,2a]
\end{array}
\right.,
\end{equation}
which, plugged into Eq.~\eqref{betas}, gives
\begin{equation} \beta_{\rm G}(h,a)\approx
\left\{
\begin{array}{ll}
\frac{L+1}{L} \frac{-2 a^{-2} h}{4-(a^{-1} h)^2}\approx \beta_{\rm B}(h,a), & h \in [-2 a,\, 0)  \\[0.5em]
0, & h \in (0,\,2a]
\end{array}
\right..
\nonumber
\end{equation}
where the subscripts in the inverse microcanonical temperatures refer to the Gibbs and Boltzmann pictures.
Entirely similar results can be obtained numerically for higher dimensions or for the quantum case considered in the inset of Fig.~\ref{BEntropy}.

\section{Bose-Einstein condensation}
\label{BECGC}
As we mention in Sec.~\ref{ENSS}, the calculation of the grand partition function, Eq.~\eqref{ZGC},  for the non-interacting version of the lattice model in Eq.~\eqref{H} can be easily carried out analytically. The result is
\begin{equation}
{\cal Q} =  \prod_\qq \frac{\pi}{\beta(\ve_\qq-\mu)}
\end{equation}
which gives rise to the average occupation distribution in Eq.~\eqref{nq}. Despite the function in Eq.~\eqref{nq} is not the standard Bose-Einstein distribution, but rather its classical limit\footnote{See note \ref{classicaln}.}, it still gives rise to condensation. The first of Eqs.~\eqref{ah} and Eq.~\eqref{nq} can be used to find the chemical potential $\mu(\beta,a)$ corresponding to a given choice of the particle density $a$  and inverse temperature $\beta$. Plugging the result into Eq.~\eqref{nq} gives the average occupation of each single-particle state. Note that the chosen $\beta$ can have either sign and, in view of the fact that $n_\qq \geq 0$, it must be 
$\mu < \min_\qq \ve_\qq$ for $\beta>0$ and $\mu > \max_\qq \ve_\qq$ for $\beta<0$. This discontinuity in $\mu$ does not necessarily mean that the system undergoes a phase transition at $\beta=0$, as argued in Ref.~\cite{Rasmussen_PRL_84_3740}. Indeed, $\lim_{\beta\to0_\pm}\mu(\beta,a) = \mp\infty$, so that  $\lim_{\beta\to0_\pm} -\beta \mu(\beta,a) = a^{-1}$. Thus the grand partition function ${\cal Q} = \pi^L \prod_\qq [\beta(\ve_\qq-\mu)]^{-1}$ is not singular for $\beta=0$. The same is true in the presence of  a non pathological interaction term, such as $u_1$, as it is clear e.g. from the results of the transfer-matrix approach in Fig.~\ref{transmat} \cite{extended}. 

\begin{figure}
\includegraphics[width=5cm]{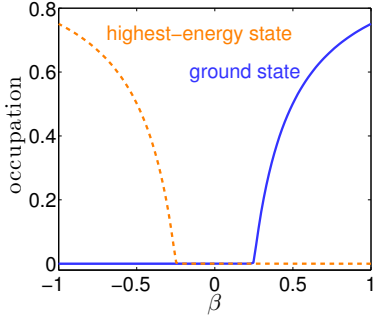}
\caption{\label{BEC3DU0} Occupation of the ground- and highest-energy- state for a $64\times64\times64$ noninteracting lattice model with $a=1$, as provided by a grand-canonical calculation.}
\end{figure}

The above sketched calculation can be easily carried out numerically. Figure~\ref{BEC3DU0} shows the behavior of the relative occupation of the extremal (kinetic) modes of a non-interacting discrete model on a 3D lattice, as the inverse temperature ranges from negative to positive temperatures. Expectedly, there exists a critical value $\beta_{\rm C}>0$ above which the ground-state of the system is macroscopically occupied. As we discussed in Sec.~\ref{PTS}, owing to the symmetry of the energy spectrum, the highest-energy state is macroscopically occupied for $\beta<-\beta_{\rm C}$.

In the presence of nonlinear interactions, only one of the extremal state maintains its extended character, while the other becomes localized.
Since it corresponds to the breaking of a continuous (phase) symmetry, the condensation into the extended state occurs at finite $\beta$ only for $d>2$, in agreement with the Mermin-Wagner theorem. 
As demonstrated in panel b) of Fig.~\ref{2Dsat}, for saturable\footnote{The analysis of the energy and temperature region in the vicinity of the extremal localized state is problematic in the case of the standard nonlinear term $u_2$, because of the pathological scaling of the energy densities.} nonlinearities the {\it self-trapping} transition occurs at finite $\beta$ also for $d<3$. 
 
 The Bogoliubov approach allows the analysis of the condensation transition at finite interactions. In the following we sketch the simplest and most standard case, i.e. the condensation into the uniform ground state for defocusing nonlinearity and arbitrary interaction term. A more detailed and general calculation, including the stability of the excited plane-wave solutions,  can be found elsewhere \cite{extended}.  As usual, in view of the decoupling of quasimomenta expected in a uniform system, we consider a perturbation of the ground-state of the form
\begin{align}
z_\rr(t) =& e^{-i \ve_{\0} t}\Big\{\sqrt{a_\beta} + \frac{1}{\sqrt{V}} \sum_{\qq \neq \0 } \Big[b_\qq\, \su_\qq e^{i (\qq\cdot \rr-\varpi_\qq t)} \nonumber\\
\label{Bog1}
   &+ b_\qq^*\, \sv_\qq^* e^{-i (\qq\cdot \rr-\varpi_\qq t)}\Big]\Big\},
\end{align}
and plug it into the equation of motion, retaining only the linear terms in $\su_\qq$ and $\sv_\qq$. 
The branch of the resulting excitation spectrum fulfilling the expected normalization relation, $|\su_\qq|^2-|\sv_\qq|^2=1$, corresponds to 
 \begin{equation}
\label{wBog1} 
\varpi_\qq =  \sqrt{(\ve_\qq-\ve_\0) \left[(\ve_\qq-\ve_\0)+2 U a_\beta u_{j}''(a_\beta)\right]}
 \end{equation}
and 
\begin{equation}
\label{uqq}
\su_\qq = \frac{U a_\beta u_{j}''(a_\beta)}{\sqrt{2 \varpi_\qq\left[U a_\beta u_{j}''(a_\beta)+\ve_\qq-\ve_\0-\varpi_\qq\right]}}, 
\end{equation}

When the Bogoliubov approximation applies, the interaction strength $U$ is incorporated into the spectrum in Eq.~\eqref{wBog1}, and the system is governed by the effectively free Hamiltonian $H_{\rm B}~=~\sum_{\qq} \varpi_\qq |b_\qq|^2$. One therefore expects that using $\varpi_\qq$ in place of $\ve_\qq$ in  Eq.~\eqref{nq}, the average occupation of the Bogoliubov modes is obtained, i.e. $n_\qq = \langle |b_\qq|^2\rangle$. A procedure similar to the one sketched above for the noninteracting case allows to study the population $a_\beta$ of the macroscopically occupied (ground) state. This is how the solid curve in Fig.~\ref{cond3DC} has been obtained. Note that when $a_\beta \approx 0$ the Bogoliubov spectrum coincides with the single-particle spectrum.

Fourier transforming the perturbed state in Eq.~\eqref{Bog1} we get 
\begin{align}
{\tilde z}_\qq &= \frac{1}{\sqrt{V}} \sum_\rr e^{i \rr \cdot \qq} z_\rr \nonumber \\
&= b_\qq {\cal U}_\qq e^{-i(\ve_\0+\varpi_\qq)t}+b_{-\qq}^* {\cal V}_{-\qq}^* e^{-i(\ve_\0-\varpi_{-\qq})t}
\end{align}
and
\begin{align}
\langle |{\tilde z}_\qq|^2 \rangle= \langle |b_\qq|^2 \rangle \left( |{\cal U}_\qq|^2 +|{\cal V}_{\qq}|^2 \right)
\end{align}
where we assumed that the time average of the terms containing the phase factors cancel out and that $\langle |b_\qq|^2 \rangle = \langle |b_{-\qq}|^2 \rangle$, on account that $\varpi_\qq = \varpi_{-\qq}$.

\section{BKT transition}
\label{aBKT}
On 2D systems the BKT transition is signalled by a change in the decay properties of the radial correlations. Denoting $C_{\rr\,\rr'} = \langle z_{\rr^{}}^{} z_{\rr'}^* \rangle $, in the defocusing case one expects a power law decay, $C_{\rr\,\rr'}  \sim |\rr - \rr'|^{-\alpha_\beta}$, for $\beta>\beta_{\rm BKT}>0$  and an exponential decay, $C_{\rr\,\rr'} \sim e^{-|\rr - \rr'|/\xi_\beta}$, for $\beta<\beta_{\rm BKT}$, with the decay exponent tending to $\alpha_{\beta} = \frac{1}{4}$ as the critical point is approached \cite{Small_PRA_83_013806,Polkovnikov_PNAS_103_6125}. The quantity $\xi_\beta$ is a temperature-dependent correlation length.
 The quantity   
\begin{equation}
\label{BKTc}
{\cal A}_\Omega(\beta) = \int_{|\rr -\rr'|<\sqrt{\Omega}} d\rr\,d\rr' |C_{\rr\,\rr'}|^2\sim \Omega^{1+\sigma_\beta}
\end{equation}
where
\begin{equation}
\sigma_\beta = 
\left\{ 
\begin{array}{ll}
1-\alpha_\beta & \beta>\beta_{\rm BKT}\\[0.5 em]
0 & \beta<\beta_{\rm BKT}
\end{array}
\right. 
\nonumber
\end{equation}
can be used as an indicator for the transition  \cite{Small_PRA_83_013806}. 

In view of the square modulus in the integrand of Eq.~\ref{BKTc} one expect an entirely similar behavior in the self-focusing case, 
\begin{equation}
\sigma_\beta = 
\left\{ 
\begin{array}{ll}
1-\alpha_\beta & \beta<\beta_{\rm BKT}'\\[0.5 em]
0 & \beta>\beta_{\rm BKT}'
\end{array}
\right. 
\nonumber
\end{equation}
where $\beta_{\rm BKT}'=-\beta_{\rm BKT}$.

This indeed is what we obtain in Fig.~\ref{2Dsat} a) for  self-focusing saturable interactions. Note in  particular that the  exponent at the crossing point of the  curves corresponding to different sizes is very close to the expected value  $\sigma_\beta=3/4$. 

We analyzed also the defocusing standard nonlinearity considered in Ref.~\cite{Small_PRA_83_013806}, qualitatively confirming the findings therein discussed \cite{extended}. We  recall that  the temperature $\beta^{-1}$ is expected to diverge as the energy density approaches the upper bound of the positive-temperature region. We observe that, conversely, the  effective temperature defined in Ref.~\cite{Small_PRA_83_013806}, $T_{\rm Small} = 2 h + a \kappa -[2 U u_j(a) - 8 a]$ tends to a finite value. For the considered standard nonlinearity we get $T_{\rm Small} = 2 U a^2 -( U a^2 - 8 a) = U a^2 + 8 a$.
 \section{Thermalization and Thermometry}
\label{TTA}
On sufficiently ergodic systems, the microcanonical (Boltzmann) inverse temperature can be obtained as the time average of a suitable function of the dynamical variables. When the only first integral of the motion is the total energy, such function is related to the curvatures of the ``energy sheet'' involved in the dynamics \cite{Rugh_PRL_78_772}. Generalizing this approach to equations having one further first integral  \cite{Franzosi_JSP_143_824}, the instantaneous microcanonical inverse temperature is obtained as
\begin{equation}
\label{betat}
\beta(t)= \frac{\|\nn\!\wedge\!\hh\|}{{\nabla}\! \cdot\! {\bar \vv}} \left[\nabla\! \cdot\! \left(\frac{\bar \vv}{\|\nn\!\wedge\!\hh\|}\right) - \frac{\bar \nn \cdot (\bar\nn\cdot \nabla )\bar \vv}{\|\nn\!\wedge\!\hh\|} \right]
\end{equation}
where $\nabla$ is  the gradient in the $2 V$-dimensional Euclidean space of the real and imaginary parts of the complex dynamical variables $z_\qq$ and the boldface variables are vectors in the same space. Specifically 
\begin{equation}
\vv = \bar \hh-(\bar \hh \cdot \bar \nn) \bar \nn,\qquad \hh = \nabla H,\qquad \nn = \nabla N
\end{equation}
where $H$ and $N$ are defined in Eqs.~\eqref{H} and \eqref{N}, respectively, and an overbar denotes a versor, i.e.  $\bar\nn = \nn /\|\nn\|$ and $\bar\vv = \vv /\|\vv\|$($\|\cdot \|$ is the standard Euclidean norm). Clearly, all the quantities in the r.h.s. of Eq.~\eqref{betat} depend on the instantaneous value of the field, so that $\beta(t) = \beta(\{z_\qq(t)\})$. This approach can be further generalized to equations having more than one additional first integral \cite{Franzosi_PRE_85_050101}. See Refs.~\cite{Rugh_PRE_64_055101,Davis_PRA_66_053618} for similar approaches.

\begin{figure}
\begin{tabular}{cc}
\includegraphics[height=3.cm]{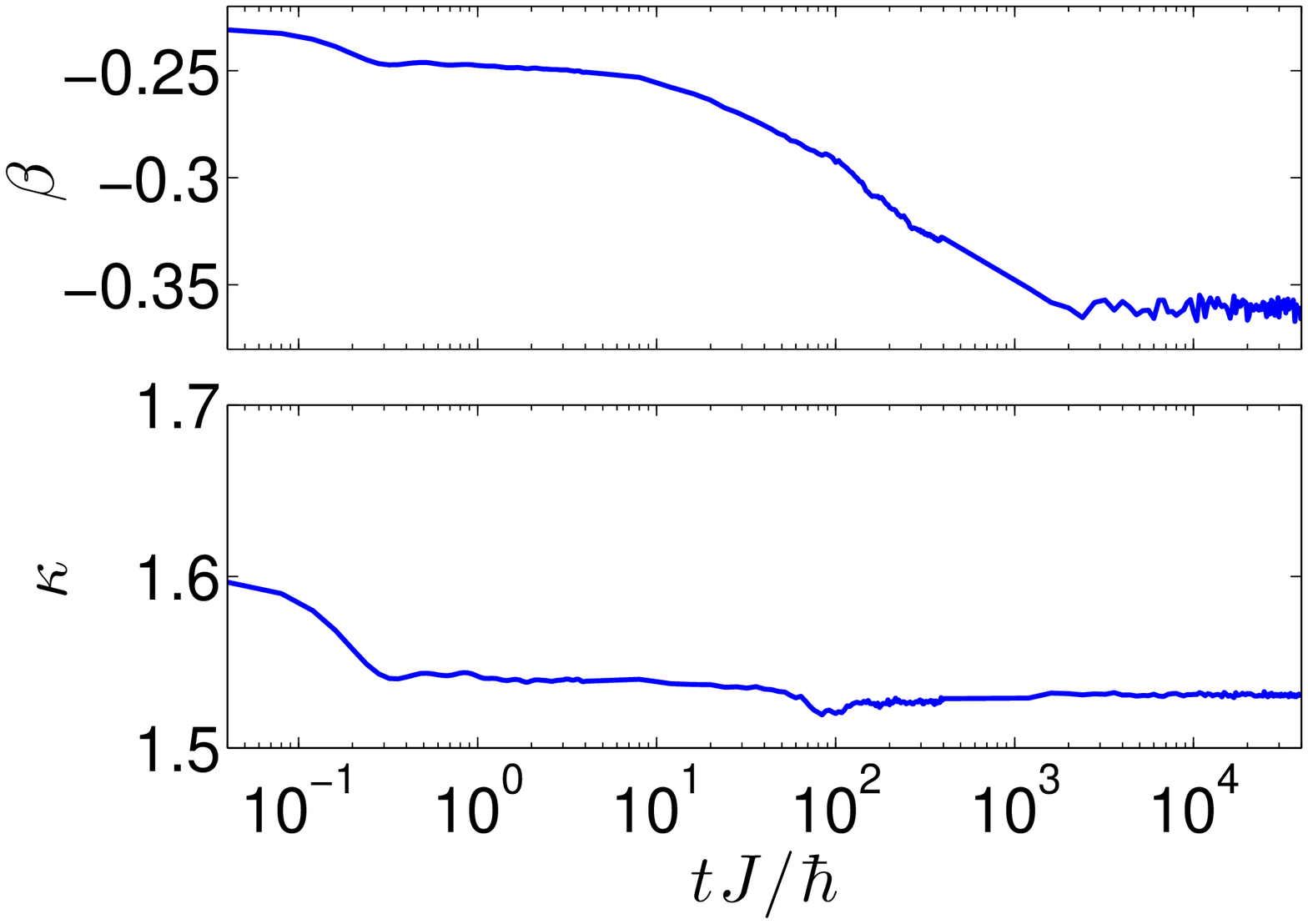} &
\includegraphics[height=3.cm]{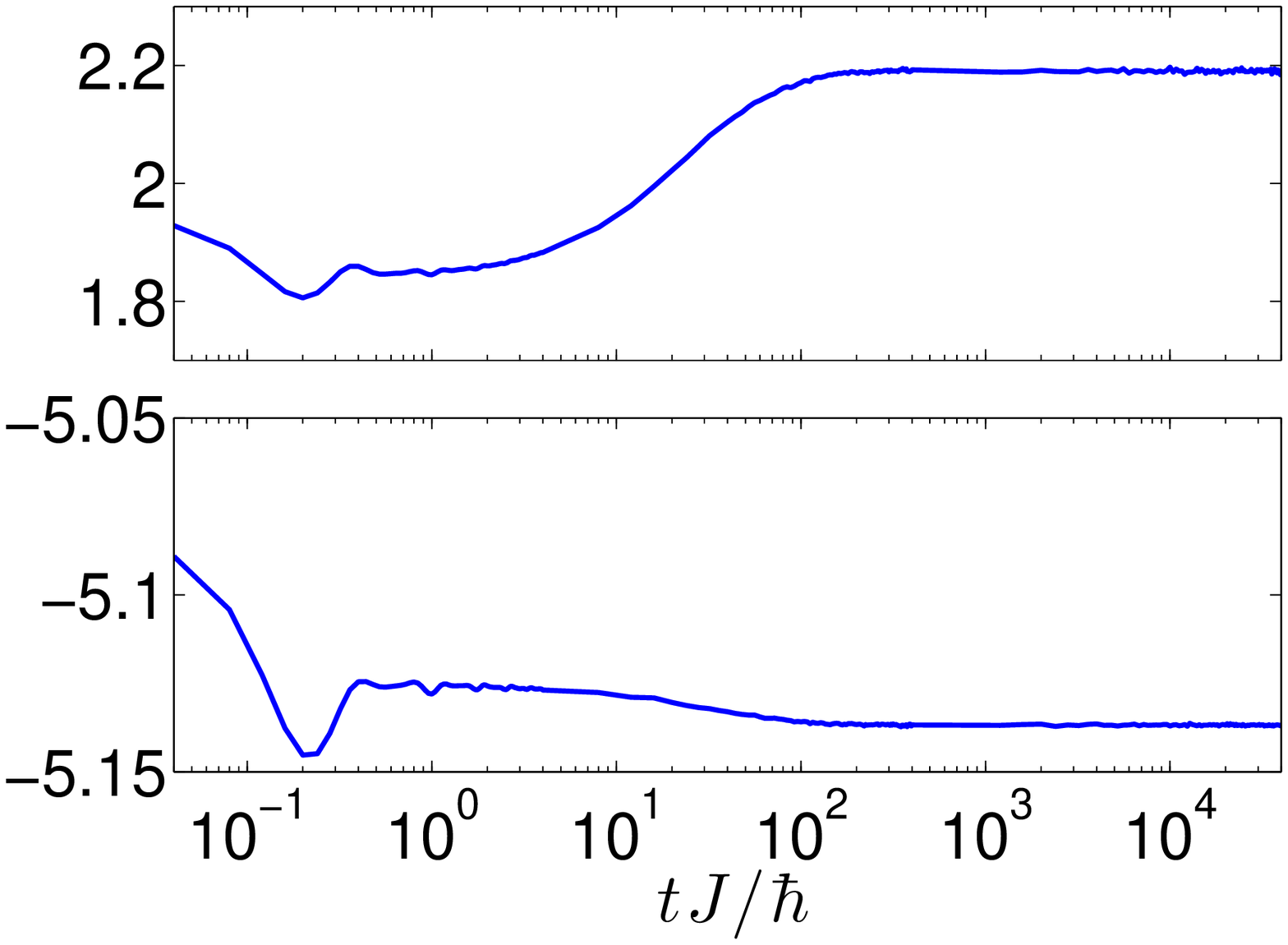}
\end{tabular}
\caption{\label{transient} Instantaneous value of the microcanonical inverse temperature and kinetic energy per particle. Left: $128 \times 128$ lattice with saturable nonlinearity  at a negative temperature. Fig.~\ref{betaC} a) has been obtained by time averaging the same data in the time window $4\times 10^4<t J\hbar< 8\times 10^4$. Right: $64 \times 64 \times 64$ lattice with standard nonlinearity at a positive temperature. The rightmost green circle in Fig.~\ref{cond3DC} has been obtained by time averaging the same data in the time window $4\times 10^4<t J \hbar< 8\times 10^4$. In both cases $U=0.75$, $a=1$.
}
\end{figure}

As we mention in Sec.~\ref{TTS}, we find that the dynamics dictated by the lattice Hamiltonian in Eq.~\eqref{H} brings the system to an asymptotic equilibrium state, characterized by well-defined values of observables such as the interaction or kinetic energy per particle, or the above-described instantaneous microcanonical temperature.
Specifically, we observe that, after a transient whose duration depends on the initial state and the Hamiltonian parameters, the instantaneous value of said observables oscillates about an asymptotic value. This allows the definition of time averages as in Eq.~\eqref{tavg}. Fig.~\ref{transient} shows some instances of the described equilibration process.

 Typically, we initialize the dynamics on a suitably perturbed ``plane-wave'' state, $z_\rr = Z (\sqrt{a}+\eta_{\delta} \delta_\rr) e^{i (\rr \cdot \qq+  \pi \eta_{\varphi} \varphi_\rr)}$, where $\delta_\rr$ and $\varphi_\rr$ are random numbers uniformly chosen in $[-1, 1]$, $\eta_{\delta}$ and $\eta_{\varphi}$ control the magnitude of the random perturbations and $Z$  is a normalization constant enforcing the desired particle density. It should be noted that unperturbed ``plane wave'' states can be either linearly stable or unstable depending on $\qq$ \cite{Smerzi_PRL_89_170402,extended}. Plane-waves having an energy close to that of the ground-state are typically stable, and hence the relevant dynamics can be non ergodic. We checked that a suitable amount of noise destroys stability, so that equilibrium can be reached also for small energies. This may require long equilibration times. Conversely, for unstable plane-wave modes, a vanishingly small noise is sufficient to trigger a modulational instability that drives the system away from the initial state very quickly. We remark that these modulationally unstable states do not necessarily end up having a negative microcanonical temperature. Actually, for suitably large densities (or interaction strengths), the whole band of ``plane-wave'' modes can give rise  to positive-temperature asymptotic states \cite{Rasmussen_PRL_84_3740,extended}.

\begin{figure}
\begin{tabular}{cc}
\includegraphics[height=4.4cm]{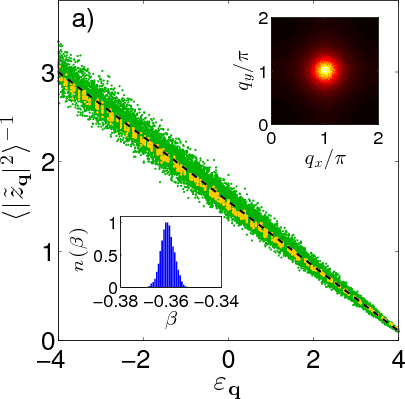} &
\includegraphics[height=4.4cm]{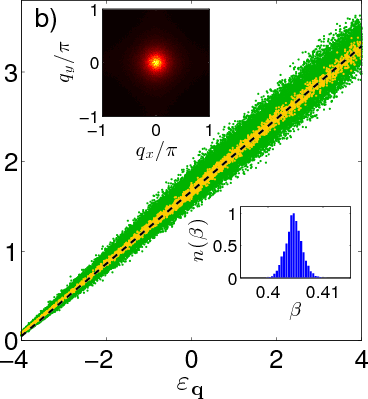}
\end{tabular}
\caption{\label{betaC} Time-averaged distribution for the occupation of the single-particle modes in equilibrium states at small $|\beta|$. The green dots refer to the whole lattice, while the yellow dots refer to a sublattice whose volume is $1/16$ of the whole lattice; the slope of the dashed black (straight) line is the time-averaged value of the instantaneous microcanonical inverse temperature. The density plots in the upper insets show the average distribution according to quasimomentum. The lower insets contain histograms of the values assumed by the instantaneous inverse temperature in the considered time window of $\Delta t= 4\times 10^4 \hbar J^{-1}$. In both cases $a=1$, $U=0.75$, and the nonlinearity is of the saturable kind. Panels a) and b) refer to two-dimensional $128 \times 128$ and $256 \times 256$ lattices, respectively.}
\end{figure}

\begin{figure}
\includegraphics[height=4.4cm]{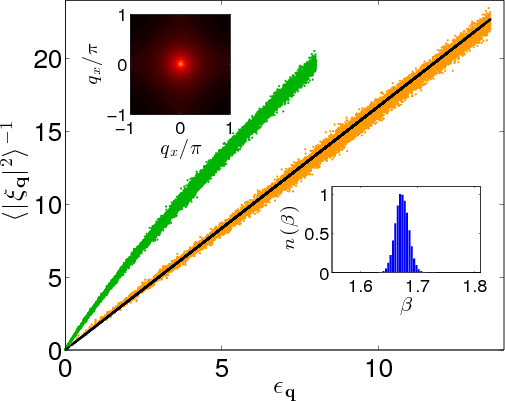} 
\caption{\label{BogM} Time-averaged distribution for the mode occupation in a $128 \times 128$ lattice with standard nonlinearity, $U=10.0$, $a=1.0$. Green dots refer to the single-particle modes, i.e. $\xi_\qq = {\tilde z}_\qq$ and  $\epsilon_\qq = \varepsilon_\qq - \varepsilon_\0$. We translated the energy spectrum for better comparison with the orange dots, which refer to the Bogoliubov quasiparticle modes, i.e. $\xi_\qq = b_\qq$ and  $\epsilon_\qq = \varpi_\qq$. The slope of the black straight corresponds to the average microcanonical temperature. The insets are obtained as in Fig.~\ref{betaC}, except that we used a logarithmic scale in the density plot of the quasimomentum distribution.}
\end{figure}

As we mention in Sec.~\ref{TTS} the asymptotic average occupation distribution of the relevant modes of the dynamics provides a further proof that the system has reached equilibrium.  This is demonstrated in Figs.~\ref{betaC} and \ref{BogM}.  The  histograms in the lower insets show that the instantaneous microcanonical temperature performs small oscillations around an asymptotic value, while the density plot in the upper insets show the quasimomentum average distribution. The scatter plots in the main figure show the average occupation of the lattice modes.
In Fig.~\ref{betaC} the temperature is comparatively large, and neither extremal state is macroscopically occupied. Therefore, the prediction of Eq.~\eqref{nq} is fulfilled by single-particle states. As we discuss in Ref.~\ref{TTS}, it is as if the interaction term simply acts as a heat bath maintaining the temperature  non-interacting system. Note that the prediction of Eq.~\eqref{nq}  applies for both positive and negative microcanonical temperatures. 

Fig.~\ref{BogM} illustrates a case where Eq.~\eqref{nq} seems to fail. It refers to a two-dimensional lattice with strong defocusing nonlinearity, $U=10.0$. The green scatter plot, obtained as in Fig.~\ref{betaC}, shows that the  occupation of the single particle modes is  ``larger than it should'' at small energies, and bends towards the expected slope ---i.e. the slope of the straight black line--- only at high energies.
This is because the system is in the condensed phase. Specifically, the average density is $a=1$, while the condensate fraction is $a_\beta = \langle|{\tilde z}_\0|^2\rangle /(a V)\approx 3/4 $.  As apparent from the orange scatter plot, the prediction of  Eq.~\eqref{nq} is recovered when the Bogoliubov modes are considered. The same results are obtained for smaller interaction strengths, although the difference between the two scatter plots is not as dramatic as in the case presented in Fig.~\ref{BogM}.

We observe that the scatter plot of $\langle|{\tilde z}_\qq|^2\rangle^{-1}$ versus $\ve_\qq$ deviates from a straight line also when the system condense into a localized state (not shown). The deviation is significant for energies close to that of the extremal localized state, while the scatter plot matches the expected linear behavior at the opposite end of the spectrum \cite{extended}. We expect that the linear behavior can be recovered on the whole energy spectrum if Bogoliubov modes are used instead of single-particle modes. However, owing to the localized character of the extremal state, the Bogoliubov approach is significantly more involved in this case.
The localized character of the dynamical state might pose some problem with respect to the thermalization of the whole lattice and its subsystems. It is indeed clear that a large number of sublattices can be found which do not feature a localized density peak. In fact, most of the sublattices would have an average particle density much smaller than that of the whole system. We verified that the average distribution for the mode occupation agrees with Eq.~\eqref{nq} when calculated in a sublattice {\em not} containing the density peak \cite{extended}.

Finally, we checked that generalizing the grand-canonical Langevin approach introduced in Ref.~\cite{Iubini_JSM_P08017},  where $\beta$ is an external parameter, produces results in agreement with the prediction of Eq.~\eqref{nq} for both positive and negative temperatures \cite{extended}.
This is one further evidence of the equivalence of the micronanonical and grand canonical ensemble for  both signs of the temperature.
 
\clearpage
\bibliography{/Users/pfb/Documents/LATEX/BIBTEX/general,./notes}

\end{document}